\documentclass[longbibliography,twocolumn,showpacs]{revtex4-1}
\usepackage{graphicx}
\usepackage[usenames,dvipsnames,svgnames,table]{xcolor}
\usepackage{amsmath,amssymb,latexsym,graphics,epsfig}
\usepackage{pstricks}
\usepackage{cancel}

\newcommand \be {\begin{equation}}
\newcommand \ee {\end{equation}}
\newcommand \bea {\begin{eqnarray}}
\newcommand \eea {\end{eqnarray}}

\begin{document}
\title{On self similarity and coarsening rate of a convecting bicontinuous  phase separating mixture~: effect of the viscosity contrast}
\author{Herv\'e Henry$^1$ and Gy\"orgy Tegze$^2$}
\affiliation{$^1$
 Laboratoire de Physique de la Mati\`ere Condensée, \'Ecole Polytechnique, CNRS, Universit\'e Paris-Saclay, 91128 Palaiseau  Cedex, France
}
\affiliation{$^2$Wigner Research Centre for Physics, P.O. Box 49, H-1525 Budapest, Hungary}
\date{\today}
\begin{abstract}

We present a computational study of the hydrodynamic coarsening in 3D of a critical mixture using the Cahn-Hilliard/Navier-Stokes model. The topology of the resulting intricate bicontinuous microstructure is analyzed through the principal curvatures to prove self-similar morphological evolution. We find that the self similarity exists for both systems:  isoviscous and with variable viscosity. However the two system have distinct topological character. Moreover an effective viscosity that accurately predicts coarsening rate is proposed.

\end{abstract}
\maketitle

\section{Introduction}
 
\begin{figure}
   \includegraphics[width=0.5\textwidth]{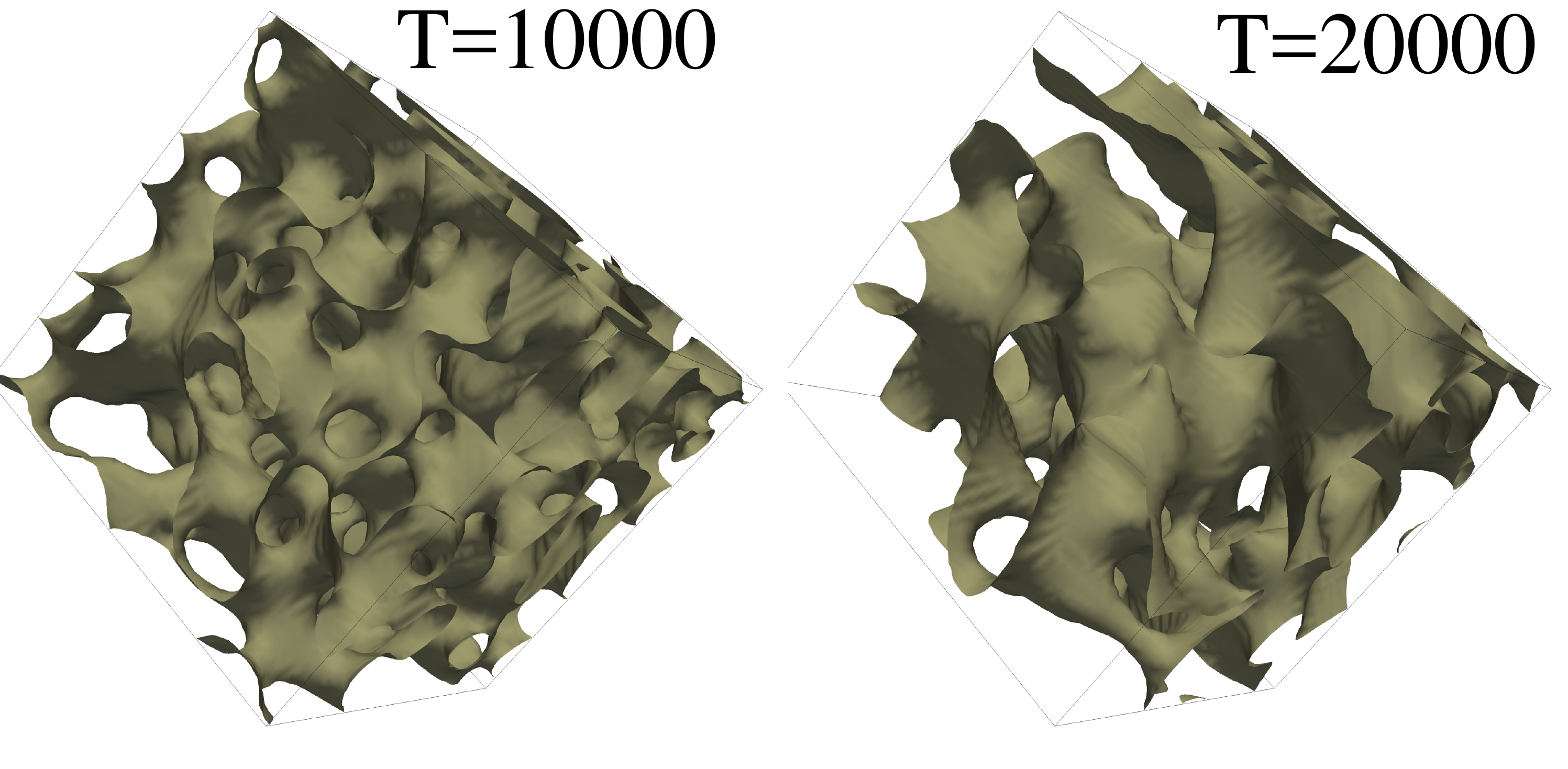}
   \caption{\label{fig_snapshots} Perspective view of the
   isosurface $c=0.5$ at different times of a  simulation. Volume fraction is $\varphi=0.5$, {the viscosity contrast between the phases is 128} and the viscosity is $\nu=2$.}
 \end{figure}

  Among the physical process leading to the formation of a microstructure, phase separation  is ubiquitous. 
It is  seen in glasses\cite{Craievich1986} and polymer blends\cite{Kumar_1996_PRL}, and can be divided into
two stages. First, the unstable mixture phase separates
at a characteristic length-scale, $l$\cite{Cahn1965}. When the volume fraction of one
phase  is close
to 0.5, the initial microstructure that arises  consists of two interlaced
percolating clusters (similar to the one presented  fig:\ref{fig_snapshots})while for significantly lower volume fractions it consists of isolated
droplets  in a matrix of the majority phase. This pattern evolves under the effect of
diffusion\cite{LSW61}  or of fluid
flow\cite{Siggia1979}, { resulting in an
increase of the characteristic length $l$ known as coarsening. It is widely
acknowledged that this  process is self similar.
In the case of diffusive 
coarsening, based either on analytical \cite{LSW61} or numerical investigations\cite{Kwon2007,Kwon2009},
strong arguments in favour of this hypothesis can be found.}

In the case of viscous coarsening,  such arguments are still
lacking. Indeed  Siggia\cite{Siggia1979},{
assuming \textit{a priori} self similarity,}  proposed using a scaling
argument, that after an initial diffusive coarsening stage where the characteristic
length grows as $t^{1/3}$, coarsening is governed by viscous fluid flow. 
This later  regime is characterized by a growth of the characteristic length at a constant rate. 
Later, { some consequences  of self-similarity were observed in}  experiments \cite{expe1,expe2} and numerical
simulations\cite{Appert1995visc,Bastea97visc,Kendon2001inertial} that 
were conducted at symmetric composition where phases share the same viscosity.
In further investigations, laws that account for inertial 
effects were also proposed\cite{Furukawainertial,Juryscaling,Elderinertial,Kendon2000scaling,inertialcheck}. 
{However these were limited to the symmetric case and
the existence of the viscous  self similar coarsening regime remains to be
uncovered  when the symmetry of the composition, or the kinetics is
broken. Moreover, the method of analysis was based on analyzing the structure functions, that
is loosing accuracy at low wavenumbers, and  more importantly it gives
no direct information about the topology of the microstructure.  } 

Herein, inspired by
recent X-Ray tomography\cite{Bouttes2014,Bouttes2016} experiments,
we explore the effect of viscosity contrast of
the phases on {the persistence of self similarity} and the
topology of the  interconnected structure.
Our analysis bases on simulations using the Cahn-Hilliard/Navier-Stokes (NSCH) model, {and characterizing the geometrical features of the microstructure using recent advanced methods \cite{Kwon2007,Araki2000}}

We first
study the hydrodynamic coarsening of an isoviscous sample as reference, { and discuss the domain of validity of Siggia's scaling}. 
Next  the effects of {kinetic symmetry breaking} (viscosity contrast) are considered and  quantitative measures to the changes of the microstructure are given.

\section{Theory and modeling}
{The thermodynamics of a binary fluid is well described by the diffuse interface theory of Cahn and Hilliard\cite{Cahn1958}.
The simplest symmetric form of the Cahn-Hilliard free-energy reads as:
\begin{equation}
  \mathcal F = \int \epsilon^2 (\nabla c)^2 + A (c^2(c-1)^2)
\end{equation}
}
{Here $\epsilon$ and $A$ are model parameters, that are used to adjust interface tension to $\gamma=0.0042$.} as in  reference\cite{Kendon2001inertial}.
 The coarsening dynamics via convection and diffusion is governed by the coupled Navier-Stokes  (eq.\ref{eq:NS2}) and the convective Cahn-Hilliard (CH) (eq.\ref{eq:CH}) equations (NSCH){, also known as model H \cite{Hohenberg_RMP}.
Thermal fluctuations, were neglected assuming they  are small on the characteristic scale of the microstructure.
}
The Navier-Stokes/Cahn-Hilliard\cite{Anderson1997} model was used along with the incompressibility constraint (eq.\ref{eq:PE}):
\begin{eqnarray}
	\partial_t c + \mathbf{v} \cdot \nabla c &=& -D \triangle \mu \label{eq:CH}\\
	\partial_t \mathbf{v} + \nabla \cdot (\mathbf{v} \otimes \mathbf{v}) &=& { \frac{-1}{\rho}(\nabla p + c \nabla\mu)} \label{eq:NS1}\\
	&+&  \nabla \cdot \left(\frac{\nu{(c)}}{2}(\nabla\mathbf{v}+\nabla\mathbf{v}^T)\right)\label{eq:NS2} \nonumber\\
\nabla\cdot \mathbf{v} &=& 0 \label{eq:PE}
\end{eqnarray}
In the Cahn-Hilliard equation (Eq. \ref{eq:CH}),  $D$ is the diffusion constant, $\mu=\delta \mathcal F / \delta c$ is the chemical potential that derives from the CH free energy. 
{
In the Navier-Stokes equation (Eq. \ref{eq:NS1})
the  $-\nabla p$ term on the RHS includes a Lagrangian multiplier that forces incompressibility.
The second  term  is the thermodynamic stress,  and accounts for capillary forces.  
The last term accounts for the viscous dissipation, $\nu{(c)}=\nu_h (1-c)+\nu_l(c)$ is the composition dependent kinematic viscosity.
$\rho$ is the mass density and was chosen to unity except specified otherwise.
We define here the viscosity contrast as the ratio of the high and low viscosity of the species ($VC=\nu_h/\nu_l$).
While, according to the Stokes-Einstein relation, varying viscosity, implies
concentration dependent diffusivity, in the late stage of  coarsening one can
assume local equilibrium at the interface. Therefore simplifying to a
homogeneous diffusion equation does not  affect the coarsening of the
microstructure. In addition, the absence of viscoelastic terms is valid under
the assumption that the shear modulus is sufficiently high\cite{Tanaka2000}.
{The model equations were simulated numerically  using standard
approaches\cite{Orszag_1969_PHYSFLUIDS,Orszag_1972_PRL,Liu_2003_PHYSD,Zhu_1999_PRE} that are described in the supplementary material 
\cite{suplmat} together with a more detailed  description of the model
equations that is inspired by \cite{Barry2011,Gyula2016PRE} . The analysis of the results allowed us to extract a characteristic
length scale $l$  that is defined as the ratio between the total volume and the
total interface between the phases and other statistical quantities such as the
probability distribution function of the curvatures of the interface\cite{Goldman2005} or the
structure functions. The detailed description of the method used to compute such
quantities is also given in  the supplementary material 
\cite{suplmat}}
 
The NSCH model reproduces well the initial phase separation followed by the
coarsening of the microstructure that is due to diffusion at small length-scales
with a characteristic length-scale growing as
$l\propto t^{1/3}$\cite{LSW61,Cahn1966diffusive}. At larger length scales the coarsening
is driven by convection that is governed by surface tension and viscous
dissipation. As a result $l$ grows linearly: $l= v_0t \propto \gamma/\nu t$ \cite{Siggia1979} where $\gamma$ is the surface tension and $\nu$ is the viscosity of the fluid when $VC=1$.
The transition from the diffusive growth to a viscous growth occurs  when $v_0$ is much larger than the growth velocity associated 
with diffusion (which itself  is a function of the mobility of chemical species). This translates into the fact that the Peclet number ($Pe=lv_0/D$) is large. 
Finally the viscous growth law  looses its validity   when  inertial effects
cannot  be neglected (the Reynolds number $Re$, defined as  $l/l_0$ where
$l_0=\nu^2/(\gamma\rho)$ becomes large). { Here we have limited ourselves to the
viscous coarsening of a phase separated mixture, assuming that the viscosity was
sufficiently high to avoid the effects of fluid flow  during the 
initial  phase separation and before well defined phases are present and 
coarsenning takes place\cite{Tanaka1998}. It is important to note that during
the course of the coarsening, since $l$ is growing, these two numbers grow
(proportionally to $l$). This indicates that the characteristic size of the flow
is the characteristic size of the microstructure and change with times. As a
result, during the coarsening of a bicontinuous structure, both $l$ and $Re$
will increase and there will be a transition from a diffusive coarsening regime
where $Pe<<1$ to a viscous dominated regime ($Pe>>1$ and $Re<<1$ followed by an
inertia dominated regime ($Re>>1$)\cite{Kendon1999scaling}. Here we have focused on the well defined Siggia regime for which   $Pe>>1$ and $Re<<1$.  

 }

\begin{figure}
   \includegraphics[width=0.5\textwidth]{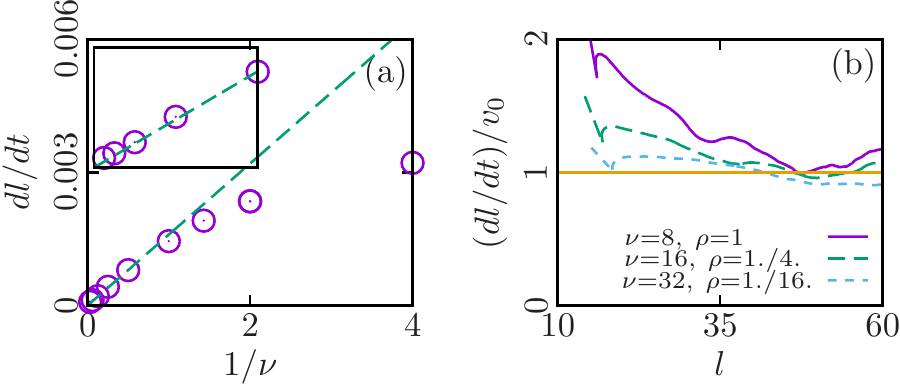}
   \caption{\label{fig_viscosity}\textbf{(a)} growth velocity as a function of viscosity
    for a volume fraction of $0.5$. In the inset a zoom on the linear regime is
    shown. \textbf{(b)} $v/v_0=dl/dt/v_0$ as a function of $l$
    for resp. $\rho=1, 0.25, 0.0625$ and resp. $\nu=8, 16, 32$ with resp. solid, long and short dash. Here $v_0$ is the value computed in the case  $\rho=0.0625,\nu=32$ (rescaled for $\rho=1, 0.25$). }
  \end{figure}

   \section{Results}
   { Since  we consider  the effect of the symmetry breaking induced by the viscosity contrast on the viscous coarsening 
  we have  chosen to limit ourselves to the case where the volume fraction of each phase is $0.5$ 
 for which the bicontinuous morphology, that is necessary to the Siggia's scaling,   
 is  more robust.  First  we present a few results in the case of the iso-viscosity regime  and
 briefly discuss the effects of diffusion and of inertia in this case. Then, the main results of
 this work, about the effects of symmetry breaking are presented. 

 \subsection{Symmetric  regime}
   This section is devoted to the determination of the parameters for which the Siggia regime is valid. Indeed, 
   while the three different regimes have been discussed at length in previous work, there is still no clear determination of   where the transition occur.  To this purpose, we first give an estimate of the diffusive effects as a function of the $Pe$ number. Then we determine the value above which the inertial terms are becoming significant.  
    
   To this purpose, we consider various parameter sets for which the Reynolds
   number and the diffusion process are kept unchanged while the Siggia flow
   rate is changed. Hence we change the value of the Peclet number without
   altering neither the relative importance of inertial terms and the absolute
   value of the diffusive contribution to coarsening thanks to the  following transformation:
   \begin{eqnarray}
     \nu& \leadsto & a\nu\\
     \rho& \leadsto & a^2\rho
   \end{eqnarray} 
  where $a$ is a real constant.  Indeed  $l_0=\nu^2/(\gamma\rho) $
   (and the Reynolds number)  is unchanged while $v_0=\gamma/(\nu \rho)$ is
   multiplied by $a$. More precisely if a given field $v(\mathbf{x},t)$ was solution of the
   Navier Stokes equation, for the original parameter set, $a v(\mathbf{x},t)$ will be a
   solution  with the transformed parameter set if the  diffusive effects  are
   neglectable.  In this situation, the coarsening
   rate with the transformed parameters will be $a$ times the coarsening rate
   with the original parameters and the relative importance of diffusive effects
   will be given by the difference between the computed solution and the predicted
   one. We have applied this approach to our system and the result is presented
   in figure  \ref{fig_viscosity}(b). The growth velocity multiplied by
   $1/v_0,\ 0.5/v_0,\ 0.25/v_0$ respectively  as a function of $l$ is
  plotted  for  $\nu=8,\  16,\ 32$ and $\rho=1,\ 0.25,\ 0.0625$ respectively
  where $v_0$ is the average value of $dl/dt$ obtained for $(\nu=32,\rho=0.0625)$. 
 If the diffusive
 effects are neglectable, one expect the curves to collapse while if diffusive
 effects are present, the difference between the curves is a measure of the
 diffusive effects. One can see that the curves obtained for the last two set of
 parameters collapse well while for $\nu=8$ there is a significant departure
 from the collapse. As a result, with   $\rho=1$ and the kinematic viscosity  
  $\nu<8$, diffusion effects can be neglected for values of $l$ larger than 50.
  This translates in terms of Peclet number into the fact that $Pe>1$ 
  ($\gamma=0.042$, $\rho=1$, $\nu=8$, $D=1\times A$ and $l=50$).

  We now turn to the effects of inertia.  To this purpose, 
  in  fig. \ref{fig_viscosity} (a) we  plot the
  growth velocity of domain size  as a function of the inverse of the  kinematic
  viscosity.  One can see that, as predicted by Siggia, that  for high values of $\nu$
  the growth  velocity is proportional to $1/\nu$ with a constant prefactor.  
  For smaller values there is a clear  departure from linear behaviour proposed in \cite{Siggia1979}. 
  The onset of this deviation occurs for  $\nu \approx 1$ and is significant for
  $\nu<0.5$, values for which the growth velocity is well described as constant
  over the length span considered here.  As a result, for these values, despite
  the apparent constant growth rate, there is a clear departure from the
  Siggia's scaling that is due to inertial effects. From a quantitative point of
  view, it occurs at $Re=1$ ($\nu=1$, $\gamma=0.042$, $\rho=1$, $l=25$). This
  value of $Re=1$ has to be compared with the one postulated by Siggia that was
  $\approx 100$ and that has been widely used since then. 

  Here  we have  characterized   the various regimes of domain growth and how they are related.
 We have also drawn a clear picture of the iso-viscous domain growth for a given
 value of the surface tension ($\gamma=0.0042$) and the fluid density
 ($\rho=1$).  For a kinematic viscosity ranging from 4 to 1, and domain sizes ranging
 from 5 to 100, the growth regime can be described as purely viscous. For lower
 values of $\nu$ (corresponding to $Re\approx 1$), a clear departure from this regime due to inertial effects can
 be seen.  For higher values of viscosity,(i.e $Pe<1.$) the contribution  of diffusion to
 viscosity can no longer be neglected.  

 }
\subsection{Symmetry breaking induced by the viscosity contrast}
  \begin{figure}
    \includegraphics[width=0.5\textwidth]{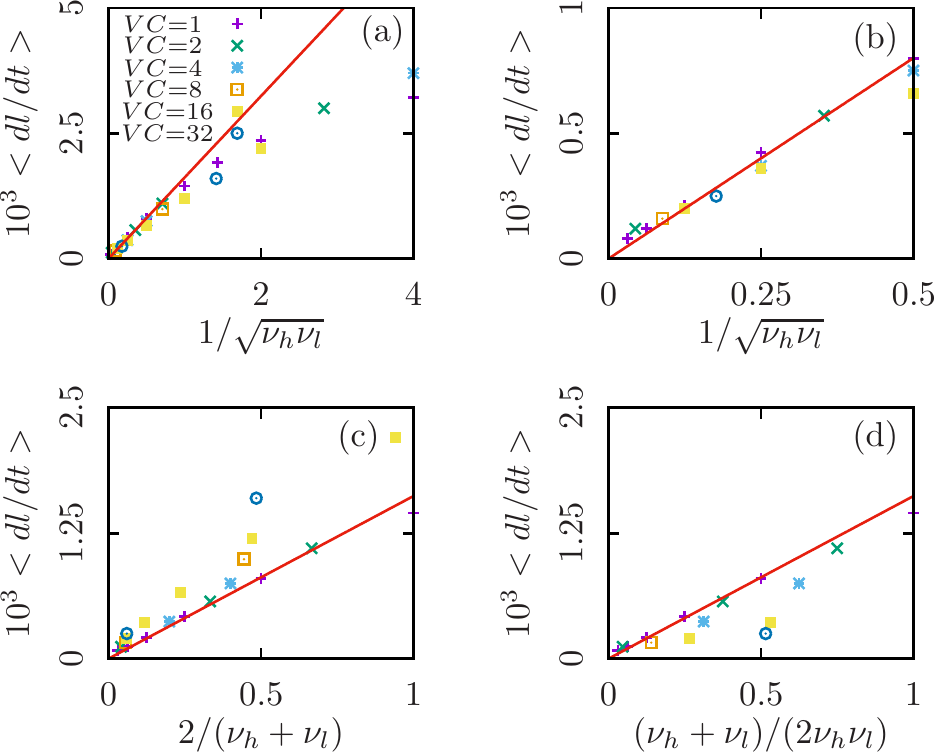}
    \caption{\label{fig:effectvisc}\textbf{(a)}: Growth rate as a function of
    $\sqrt{\nu_l\nu_h}$ for
    different values of $\nu_h$ ranging from 0.0625 to 32  and of $\nu_h/\nu_l$ equal to 2, 4, 16 ad 32. 
    The points corresponding to $\nu_h/\nu_l=1$ are the
    purple $+$.    The line  is  a guide to the eye. \textbf{(b):} same
    data with a zoom on the vicinity of the origin. \textbf{(c) (resp.
    d)} plot of the coarsening velocity versus ${\nu_{\mbox{eff}}} = (\nu_h+\nu_l)/2$ (resp.
    $1/{\nu_{\mbox{eff}}} = (1/\nu_h+1/\nu_l)/2$). }
  \end{figure}

  We now consider the evolution of the microstructure when the two  phases have
  different viscosities.  The viscosity contrast ($VC$) is the ratio
  $\nu_h/\nu_l$ and ranges from 1 to 128 in our simulations. 

  First we consider the evolution of the
  characteristic length  $l$ when  $\nu_h$ is
  small enough  to guarantee that  diffusive effects can
  be neglected.  
  In such situations,  the domain growth over 
  time   is linear with 
  a  velocity that is  a function of both $\nu_h$ and $\nu_l$. In the
  spirit of \cite{Onuki_1994_EPL} we  seek an \textit{effective
  viscosity} $\nu_{\mbox{eff}}$ for the two phase fluid, that predicts the coarsening 
  rate.
  We consider the
  following simple forms of the effective viscosity: the arithmetic mean ($
  \nu_{\mbox{eff}}=(\nu_h+\nu_l)/2)$, the geometric mean
  ($\nu_{\mbox{eff}}=\sqrt{\nu_h \nu_l}$) and  Onuki's formula ( ${1}/{\nu_{\mbox{eff}}}=({1}/{\nu_{h}}+
  {1}/{\nu_{l}})/2$) \cite{Onuki_1994_EPL}.
  The results are summarized in figure \ref{fig:effectvisc} and
  indicate that the use of the geometric mean ($\sqrt{\nu_h\nu_l}$) leads to a
  very good collapse of the curve  giving the coarsening rate as a function of
  the effective viscosity in the linear regime and still a good collapse when
  inertial effects are present. Other propositions for the effective viscosity
  are far less convincing. {Hence, the viscous growth of the
  microstructure is the same as the one that would occur if the viscosity was
  the geometric mean of the viscosities.}
   
    Finally  we describe  the effects  of the viscosity contrast on
    the microstructure itself. To this purpose we
    consider three values of the viscosity contrast (1, 16 and 128) and choose
    $\nu_h$ and $\nu_l$ so that $\sqrt{\nu_h\nu_l}=4,\ 8,\ 16 $.
    To avoid the effect of the diffusive cross over we set density as:
    $\rho=0.00625$.   Using the effective viscosity, 
    $l_0 \approx 2.10^5$ and the Péclet number is ranging from 5 to 200. Hence
    we have a set of parameters for which we expect both inertial and diffusive
    effects to be neglectable. 

    {With this parameter set   the
   Probability Distribution Functions (PDFs) 
   of the principal curvatures (rescaled by $l$) are independent of $\nu_{eff}$
   and of the initial conditions(see supplementary material \cite{suplmat}), indicating the generality of the
   results presented here}. In
   fig.\ref{fig:pdf}  the contour lines of the PDFs in the two extreme cases ($VC=1$
   and $VC=128$, with $\nu_{eff}=8$) are plotted. As expected in the   $VC=1$
   case, the PDF is
   symmetric with respect to the axis $l\kappa_1=-l\kappa_2$ and the contours
   corresponding to  two times where  $l\approx23$ and to $l\approx84$ , the  PDFs are
   indistinguishable, indicating the self similar nature of the domain growth   In addition  
   the contribution of the regions where $\kappa_1$ is
   of the same sign as $\kappa_2$ is negligible.

   The self similar behaviour holds in the case $VC=128$,{ as can be seen in the plot of the structure functions (see fig.\ref{fig:pdf} )} but the plot of  PDF of the curvatures (fig.\ref{fig:pdf}(b))
  are no longer symmetric with respect to the axis $\kappa_1=-\kappa_2$  which clearly indicates the effects of the symmetry breaking. In addition, the contribution of the regions where both $\kappa_1$ and  $\kappa_2>0$  is no 
   longer negligible :  some regions  of the interface between
   the fluids  are spherical  caps ( as seen fig.\ref{fig_snapshots} which was not the case for  $VC=1$.    
   This  is also confirmed by the plot of the PDF (fig. \ref{fig:pdf} (c))of the
   Gaussian curvature where for viscosity contrast 16 and 128, the contribution
   of the $\kappa_g>0$ part of the curve is not negligible, contrary to the $VC=1$ case.
 \begin{figure} 
  \includegraphics[width=0.5\textwidth]{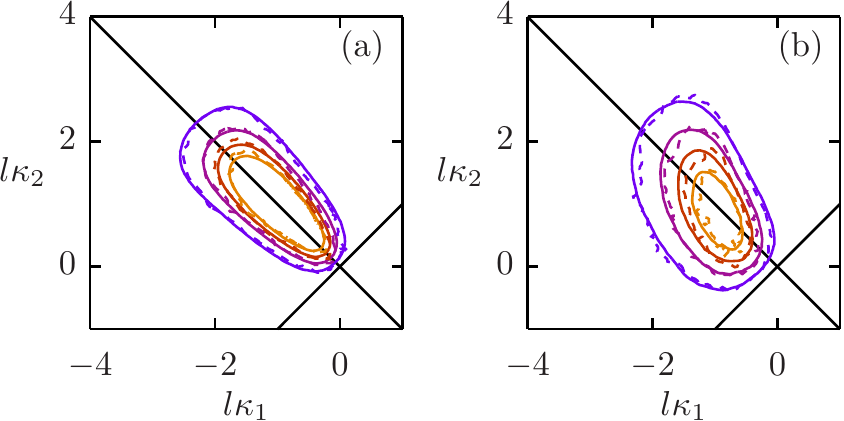}
\includegraphics[width=0.5\textwidth]{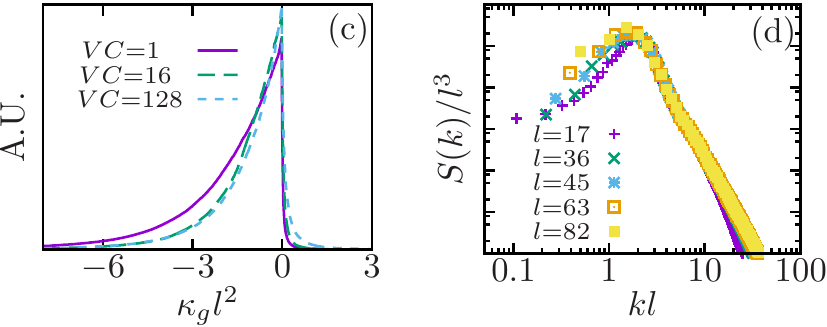}
  \caption{\label{fig:pdf} \textbf{ (a), (b)}: Contour plots of the PDF of the  principal
  curvatures rescaled by the characteristic length for two values of the
  viscosity contrast: 1 (resp. 128) taken at two times corresponding to (solid) $l\approx
  23$ (resp. 30) and to (dashed) $l\approx 88$ (resp.84. ) (dashed). The
  iso-levels are 0.0001, 0.0002, 0.0003, 0.0004 and 0.0005. \textbf{(c)}  PDF of the Gaussian curvature for three different values of the viscosity
  contrast (solid: 1, long dash 16 and short dash: 128). \textbf{(d)} Plot of the normalized structure functions taken
at different time corresponding to the value of $l$ indicated on the graph.
{The very good collapse of the curves for the lowest values of $l$  confirms the self similar nature of the coarsening
process. The low wavenumber departure from the collapse for $l=63$ and $l=82$
can be attributed to discretization effects\cite{Kendon2001inertial} }} 
\end{figure}

 Finally, we show the evolution of the rescaled genius number  ($g$, which is proportional to the rescaled mean Gaussian curvature and a simple function of the Euler's characteristic.\cite{suplmat}), and of the rescaled mean curvature $<l\kappa_m>$  as a function of $l$ for these three values 
   of $VC$.   
   {
After an initial transient,
   as expected for a self-similar growth both  the genius and the mean curvature  are  approximately constant
   for a given value of $VC$. In the case of  the genius number,  (fig. \ref{gaussmean} (a))the values computed  are similar to the one found in \cite{Kwon2010} ($\approx 0.13$) in the case of diffusive coarsening and 
   increasing $VC$  induces a decrease of $g$.  Nevertheless, the effect  is
   small and there is an 
increase in measurement error as $l$ increases.  In contrast, the effect on the  average mean curvature, (fig. \ref{gaussmean}. (b) ) are much clearer. Indeed,  for $VC=1$, it is  $0$ (up to numerical/statistical  errors)for symmetry reasons. When $VC$ is increased, there is a clear departure from 
   this value that confirms the symmetry breaking.   
    }

   {The  experimental results from  \cite{Bouttes2016,Bouttes2014} give a growth of the Euler characteristic as $0.98l^{1/3}$ for a  $VC\approx 10^5$ and a volume fraction $\approx0.45$
   while  our results for $VC=128$ would correspond to a growth as $0.7l^{1/3}$. The difference can be attributed to the dramatic difference in parameter values.  From a more qualitative point of view, 
   we find noteworthy the fact that the PDFs of Gaussian curvature when the
   viscosity contrast is increased present a significantly higher contribution
   of $\kappa_g>0$, that corresponds to spherical caps (that are absent in the
   case $VC=1$) since in  the experiments with high values of $VC$, spherical
   inclusions are observed.}

   \begin{figure} 
     \includegraphics[width=0.5\textwidth]{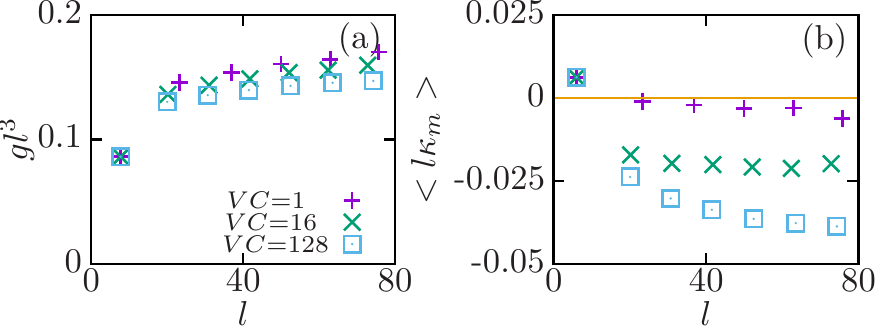}
     \caption{\label{gaussmean}{(a) Plot of the rescaled genius   as a function
     of $l$ for $VC=1$ (+), 16 (X) and 128($\Box$).   (b) Plot of the rescaled
     average mean curvature as a function of $l$ for $VC=1$ (+), 16 (X) and 128($\Box$).   }}
   \end{figure}

  \section{Conclusion} 
  {
  Here, the hydrodynamical coarsening of a  two
  phase mixture at symmetric composition is studied using constant and varying viscosity.
{Firstly, we have  challenged the assumption of self-similarity.
}
The analysis of the PDFs of the principal
  curvatures gives strong arguments in favour of the self
  similar nature of the viscous coarsening  in both cases. 
In addition, the
  analysis presented here is suitable to describe the geometry and topology
  of the microstructure. More specifically, the effects of the symmetry breaking on
  the morphology are  described  and qualitative agreement with
  experiments\cite{Bouttes2014,Bouttes2016} is found. }
  When considering the kinetics of the coarsening process, we
  show that  the linear growth regime  predicted by  Siggia\cite{Siggia1979} actually exists
  in the case where the two fluids share the same viscosity for values of the
  Reynolds number below 1. 
  When symmetry is broken by
  introducing viscosity contrast, the self similar linear growth
  still persists. {
Furthermore, our analysis allowed us to propose a formula for an effective viscosity that accurately predicts the coarsening rate of the microstructure and may be used to estimate the magnitude of flow induced coarsening in experiments,
This is in contrast with the thoroughly studied case of viscoelastic systems, where departure from self-similarity is observed\cite{Tanaka2000,Araki2000}}.

 {
   Further understanding of the
  microstructure formation during coarsening should be gained by study of the
  pattern formation process in the off critical mixture $\varphi\ne0.5$ where we
  expect to observe dramatic topological changes during the coarsening process.  
  }

 \begin{acknowledgments}
%\acknowledgements
 This work was granted access to the HPC resources of IDRIS  under the
 allocation 2016-A0022B07727 made by GENCI (Grand Équipement National
 de Calcul Intensif). The authors benefited from the support of the Chaire  St-Gobain of École Polytechnique for
 travel expenses, and was supported by projects K-115959 and KKP-126749 of the National Research, Development and Innovation Office (NKFIH), Hungary. 
And most importantly they benefited from stimulating
 discussion with  D. Vandembrouq and E. Gouillart.
 \end{acknowledgments}


\begin{thebibliography}{37}
\bibitem [{\citenamefont {Henry}\ and\ \citenamefont {Tegze}()}]{suplmat}%
  \BibitemOpen
  \bibfield  {author} {\bibinfo {author} {\bibfnamefont {Hervé}\ \bibnamefont
  {Henry}}\ and\ \bibinfo {author} {\bibfnamefont {Gyorgy}\ \bibnamefont
  {Tegze}},\ }\href@noop {} {\enquote {\bibinfo {title} {Suplemental material
  url to be given by publisher},}\ }\BibitemShut {NoStop}%
\end{thebibliography}
 \end{document}